\documentclass{article}%
\usepackage{amsmath}
\usepackage{amsfonts}
\usepackage{amssymb}
\usepackage{graphicx}
\usepackage{multirow}
\usepackage{fleqn}

\newcommand{\be}{\begin{equation}} \newcommand{\ee}{\end{equation}}
\newcommand{\ba}{\begin{eqnarray}} \newcommand{\ea}{\end{eqnarray}}
\newcommand{\ban}{\begin{eqnarray*}}
\newcommand{\ean}{\end{eqnarray*}}
\newcommand{\barr}{\be\left\{\begin{array}}
\newcommand{\earr}{\end{array}\right.\ee}
 
\def\pmb#1{\setbox0=\hbox{$#1$}\kern-.025em\copy0\kern-\wd0
\kern-0.05em\copy0\kern-\wd0\kern-.025em\raise.0233em\box0}
   
\newcommand{\half}{\mbox{$\frac{1}{2}$}}

\newcommand{\p}{\partial}

\title{Uhlenbeck's Complaint}
\author{E.A. Spiegel \\
Astronomy Department, Columbia University, New York, 10012, USA} 
                                          
\begin{document}
\maketitle

\begin{abstract}
The passage from kinetic theory to fluid dynamics as discussed by Hilbert has perplexed Uhlenbeck.  Herein, I try to smooth over the discord.
\end{abstract}

\section{The Subject} 
The invitation to participate in this commemoration of Michel H\'enon 
contained the suggestion that the topics discussed should relate to subjects that Michel had
worked on.  And, though Michel and I did not normally work in overlapping topics, there were 
several points at which our main interests were tangent.  So my choice was clear.  Everyone in this 
audience must know of Michel's decisive work in stellar dynamics.    Much of my own work 
has been in the study of fluid dynamics and therein lies a point of tangency.  
Michel has studied the dynamics of gases that are made up of stars and I have been 
preoccupied with the dynamics of gases of which stars are made.  The connection is 
well expressed by this remark of Ogorodnikov (1965):
\bigskip

{\obeylines
                                 In order to exhibit more clearly the kinematics of highly
                                 rarefied media, and of stellar systems in particular, it is
                                 useful to make a comparison with the motion of a fluid.}
 
\bigskip                                
My aim here is to describe in general terms how this connection comes about.  I want to explain
how a dynamical description of a gas of individual particles (stars if you will) in six dimensions
is reduced to the description of a continuous medium in three dimensions.  The passage is
effected by a  mathematical procedure called by various names such as ``dimensional reduction" or ``contraction."  One or another version of the procedure appears in many subjects and is often insinuated into a scientific theory with hardly a comment.  For example, we have as many neurons in our brains as there are stars in our galaxy.  How is it that the brain has (at least in the past) been regarded as a single entity that enables me to talk to you and you to hear me and (I hope) even understand what I say?  Of course, that particular reduction is still too hard for us to carry out explicitly but statistical physics is (or so I think) easier than neurobiology and so we are furnished with an example of this process that is easy enough to deal with and yet challenging enough to be interesting.  

What is significant here in the case of gases made up of stars is that they are often rarefied.  Unfortunately, the traditional method of passing from the dynamics of a gas of  discrete particles to a continuum description leads to the time-honored Navier-Stokes equations and these fare badly when we use them to describe the dynamics of a rarefied gas.  I heard Uhlenbeck complain often about this problem when I was graduate student in astronomy in Ann Arbor, Michigan sixty years ago.  He finally gave up the struggle to correct this situation and, with his students, turned to the use of kinetic theory to discuss such phenomena as the propagation of sound in rarefied gases.  But much of the inadequacy of the N-S equations in describing fluid dynamics of rarefied gases originates in the Chapman-Enskog procedure for going from the discrete kinetic theory model to the continuum description of rarefied gas dynamics.  I shall not go into great detail about that problem as it is too lengthy and has been described elsewhere in several places (including Chen (2001); Chen and Spiegel; Spiegel and Thiffeault; Dellar).  Instead I shall describe another feature of the connection between kinetic theory and gas dynamics that troubled Uhlenbeck and is the ``complaint" referred to in the title.
\section{Some Particulars}  
Imagine an enormous swarm of $N$ point masses, each of mass $m$, in a Euclidean three-dimensional space.  At any time each point mass is distinguished by its position, 
${\bf x}$, and its velocity, 
${\bf v}$, so that it may be represented by a point in a six-dimensional phase space.  A morsel of this gas of stars or particles in the tiny volume ${d\bf x}\, d{\bf v}$ is
expected to contain,  
\be
dN = {\mathcal F} ({\bf x}, {\bf v}, t) d{\bf x}\, d{\bf v} \ee
particles where ${\mathcal F}$ is the density of particles in phase space and $t$ is the time.  ${\mathcal F}$ is sometimes called the one-particle distribution and 
${\bf v} = \dot{\bf x}$ where the dot means time derivative.  I assume that this space stretches to infinity in all six of
its dimensions but I will not be fussy about what is going on out there.

As any given particle moves through the swarm of its fellows it will interact with one or more of them by means of gravitational potentials that can cause it to change its velocity.   In Boltzmann's kinetic model the interactions are usually thought of as collisions between pairs of particles mediated (if that is the word I
want) by forces with short range.  This is not the best description of stellar dynamics, though it figures in several topics, including radiative transfer.  For a stellar system, a more appropriate model is one in which the particle feels long-range gravitational interactions with many particles at once.  This is analogous to what is going on in an ionized medium, or plasma; it is characterized in the book of Clemmow and Dougherty (1969) in this way:
\bigskip

{\obeylines
          When the particles of a gas interact according to an inverse square 
          law, collisions are not predominantly binary, that is to say [the collision
          term] for a ``test particle'' at (${\bf x, v},t$) does not derive mainly
          from the possibility that other particles approach very closely and
         deflect it.  The cumulative effect of more distant, but more numerous,
         particles is more important.  Of course the influence of particles
         considerably removed  from the test particle is already included in the
         field [produced by the smoothed distribution of all the particles] and 
         [the collision term] is supposed to account only for the error in 
         neglecting the discreteness of the particles.  ...  Here it suffices to say 
         that when a test particle is subject to simultaneous ``grazing'' collisions 
         in this way its progress in velocity space becomes a random walk, 
         superposed on the ordered motion due to the macroscopic field.}

\bigskip

They go on to describe how Rosenbluth  et al. (1957) develop an approximation 
procedure to reduce Boltzmann's collision term to the more manageable Fokker-Planck form (see below) and to explain why the procedure makes physical sense.   To make this statement more meaningful it may be helpful to mention the basic equations.
\section{Technical Details}
Let us assume that the particles interact through potentials and that they behave
according to Hamiltonian dynamics, for which \be
\p_i v^i + \frac{\p \dot v^i}{\p {v^i}} = 0 \ee
where $\p_i=\p\, / \p x^i$, $\p_t = \p /\p t$, $i=1,2,3$ and summation over repeated indices is understood.   

The equation governing the evolution of ${\mathcal F}$, called a continuity equation, then takes the form \be
\p_t {\mathcal F}+ {v^i \p_i} {\mathcal F} + a^i \p_{v^i} {\mathcal F} = 
{\mathcal C}[{\mathcal F}]\, .  \label{boltz}\ee                           
The term on the right --- called the collision term --- represents the effect of interactions 
of a given particle with the ambient particles.  It gives the net rate at which the particles with a ${\bf v}$ in $d{\bf v}$ enter or leave $d{\bf v}$ and it is a measure of a transition probability in velocity space.  (${\mathcal C}$ does not operate on ${\bf x}$, which it sees only as a parameter that it does not affect directly.)  

Stellar systems are often thought of as analogous to plasmas except that,
in the case of stellar systems, the interactions are gravitational rather than electric.  As
in the case of plasmas described in the quotation from Clemmow and Dougherty, there is a natural distinction between the gravitational force arising from the smoothed background of stars and the local individual interactions between pairs of stars.  This distinction is made and used by Chandrasekhar (1949) and Lewis and Keller (1962) and others.

With the smoothed distribution denoted as $<{\mathcal F}>$, let \be
\mathcal F = <{\mathcal F}> + f\, . \ee  
Then the smoothed background mass density is\be
<\rho> = m \int <{\mathcal F}> d{\bf v} \label{meandens} \ee
and this is the source of the smoothed gravitational potential $\Phi$ given by \be
{\bf \nabla}^2 \Phi = 4\pi G <\rho> \, . \label{rho}\ee 
Then, the approximate equation for the velocities of particles between the collisions is
\be
\dot{\bf v} = \nabla \Phi = {\bf a}\, \label{a}\ee

Once the effect of the smoothed density described by $<{\mathcal F}>$ is accounted for by $\Phi$, then $f$, the part of ${\mathcal F}$ left over, accounts for the effect of the density fluctuations on the collision rate with a replacement of the collision term given by
Rosenbluth et al. (1957) as \be 
{\mathcal C}[f] =  \frac{\partial}{\partial_{v^i}}\left[-A_i \, f +
\half \frac{\partial}{\partial_{v^j}}(B_{ij}\, f)\right] . \label{FP}\ee
The ``friction'' and ``diffusion'' coefficients $A_i$ and $B_{ij}$ are naturally
functionals of $f$.  This remnant of the collision term is held responsible for the individual 
collisions at relatively close range in the spirit Boltzmann's fundamental work. If we 
leave out the terms arising from $f$ we recover the Vlasov approximation from 
which spring such important outcomes as Landau damping.  To render the treatment of these effects in the case of plasmas tractable it is usual to create
an underlying stationary state by including a (typically uniform) background of positive charge that neutralizes the plasma.  

Here, we are here interested in ``gravitational plasmas."  In fact, the cosmological term that was first introduced by Einstein into his gravitational tensor describes what may be considered as a force of repulsion as was first noted by Eddington.  It may be 
expressed as a uniform background of matter with a negative gravitational mass (Spiegel, 1998) where the density of this negative mass translates into a value for the cosmological constant.  This additional term is analogous to the term that represents the background positive charge that neutralizes the charge of the electron gas in a conventional plasma.  In 
introducing the cosmological term we do well to recall the work of Lovelock (1971) as summarized in his abstract:
  
\bigskip

{\obeylines
         The Einstein tensor $G_{ij}$ is symmetric, divergence free, and a concomitant 
         of the metric tensor $g_{ab}$ together with its first two derivatives. In this paper 
         all tensors of valency two with these properties are displayed explicitly. The 
         number of independent tensors of this type depends crucially on the dimension 
         of the space, and, in the four dimensional case, the only tensors with these 
         properties are the metric and the Einstein tensors.}

\bigskip

These matters deserve a more thorough and detailed discussion that I hope to 
provide elsewhere.  My aim here has been only to suggest in a {\it qualitative} way
how the derivation of fluid equations from kinetic theory, discussed next, have some
connection with similar reductions relating to stellar dynamics.\footnote{If you intend to apply any of the notions described herein to a gas of galaxies, you would need to allow for the mushiness of galaxies in collision.  In such collisions, galaxies transform much of their macroscopic kinetic energy into the energy in the individual
motions of their constituent stars.  Though there was a time when that 
process was often overlooked but, in fact, it may be important for understandng why
galaxies are able to merge.}
  
\section{Fluid Variables} 
In the previous section I reported that the replacement of ${\mathcal F}$ by the
smoothed distribution $<{\mathcal F}>$, with $f$ neglected, gives the
famous Vlasov approximation.  In this section, I go to the other extreme and omit 
the smoothed term $<{\mathcal F}>$ and focus instead on the localized structures
described by $f$.  The aim of making this stark separation is to bring out the relation of
the one-particle distribution to the conventional fluid variables.  The models contemplated here, are assumed to possess the desirable properties of Boltzmann's model even when
these stand-ins are used to simplify certain derivations.   In particular, I assume that any model we use to derive the gas dynamical equations satisfies the conservation condition
\be
\int \Psi^A {\mathcal C}[f] d{\bf v} = 0 \label{cons}
\ee
where $A=0,i,4$ and $i=1,2,3$ and \be
\Psi^0 = m, \ \ \Psi^i = m v^i, \ \ \Psi^4 = \half m {\bf v}^2\, . \ee  

If we add up the masses of all the particles of all velocities in $d{\bf x}$, as we did for
the smoothed density in (\ref{meandens}), we obtain the mass, $\rho d{\bf x}$, of the gas in $d{\bf x}$.  Hence we find that the mass density is \be
\rho({\bf x}, t) = \int \Psi^0 f\, d{\bf v}\, . \label{11}\ee
Similarly, we find ${\bf u}$, the velocity of the fluid morsel, from \be 
\rho u^i({\bf x}, t) = \int \Psi^i f\, d{\bf v}\, .\ee

Now let the velocity of a parcel with respect to the moving morsel --- its peculiar velocity ---
be ${\pmb \xi}$: \be {\pmb{\xi}} = {\bf v} - {\bf u}. \ee
We may then define the temperature,
$T({\bf x}, t),$ by \begin{equation}
\rho\,\Re\, T({\bf x}, t) = \frac{1}{3}m \int {\pmb{\xi}}^2 f d{\bf v} \label{14}\end{equation}
where the gas constant is $\Re = k_B/m$ and $k_B$ is Boltzmann's constant.   

We thus see how the main fluid variables, $\rho$, $T$ and ${\bf u}$ are moments of the
kinetic phase space density.  You might then wonder why these variables suffice to specify the state of a fluid.  That is the result some call Hilbert's theorem to the consternation of Uhlenbeck.
\section{The Complaint}
In a brief autobiographical memoir  G.E. Uhlenbeck (1980) wrote   
\bigskip

    {\obeylines
            In the famous book of Hilbert on integral equations there 
            is a chapter devoted to the kinetic theory of gases, in which
            Hilbert claims to have {\emph proved} from the Boltzmann equation 
            [(\ref{boltz})] that the state of a gas is completely determined if one 
            knows initially the spatial dependence of the five macroscopic 
            variables $\rho$, ${\bf u}$  and $T$.}
            [Italics Uhlenbeck's; footnotes are omitted.]
\bigskip

Uhlenbeck goes on to suggest that this statement by Hilbert, 
sometimes known as Hilbert's theorem, is not a theorem at all.
He adds that
\bigskip

 {\obeylines
       On the one hand it couldn't be true, because the initial-value        
       for the Boltzmann equation (which supposedly gives a better 
       description of the state of the gas) which requires the knowledge 
       of the initial value of the distribution function $f_1({\bf r, v}, t)$ of  which 
       $\rho$, ${\bf u}$ and $T$ are only the first five moments in ${\bf v}$. But on the other 
       hand the hydrodynamical equations surely give a causal description 
       of the motion of a fluid. Otherwise how could fluid mechanics be 
       used? Clearly one has here to do not with a theorem but with a 
       paradox, which I propose to call the Hilbert paradox.}
\bigskip

To rationalize what he called `` a paradox," Uhlenbeck described Bogoliubov's ideas about the derivation of the Boltzmann equation from Liouville's theorem.   Bogoliubov's
discussion of that derivation boils down to the so-called multi-time method of asymptotic
theory.  I suspect that those who have studied dynamical system theory will recognize
that Hilbert's discussion may be an early (a first?) `proof' of the center manifold theorem, at least in kinetic theory.  Whether there really is a theorem in this 
context, despite Uhlenbeck's doubts, depends on the conditions imposed --- in the fine
print.  There may be a theorem lurking here if the gas is spatially homogenous
and the spectrum of the linearization of ${\mathcal C}$ is discrete. I am happy to settle for
the qualitative discussion of Uhlenbeck and to hide behind this remark attributed to Feynman: "If it is true, why do I have to prove it?"  However, Uhlenbeck was right to call on Bogoliubov's ideas to rationalize Hilbert's result since Bogoliubov's ideas have been credited with inspiring the proof of the center manifold theorem.   
\section{The Gas Dynamical Equations}
In fact, the story that I am closing in on after this lengthy preamble is about the actual derivation of the equations of gas dynamics from kinetic theory, primarily from Boltzmann's equation and its imitations.   At issue is the procedure for deriving  a reduced set of equations in the three-dimensional space of our physical world, equations governing the lower order moments of the distribution function --- the fluid variables --- for the case of
a rarefied gas..  As I have mentioned, such a reduced, or fluid dynamical description, has been thought to be a useful guide to the complexities of stellar dynamics.   And what I am suggesting is that Hilbert's result is really an asymptotic conclusion.  Unfortunately, Hilbert never managed to go beyond the Euler equation of fluid dynamics though he did develop an expansion procedure for the purpose, as discussed by Ferziger and Kaper (19YY).  Hilbert's difficulty in going further caused others to seek variants on Hilbert's use of his expansion.  The outgrowth of that effort was Chapman-Enskog procedure, introduced by (separately) by Chapman and Enskog from different viewpoints (see Grad (19ZZ)).  It dominated the field for most of the past century was that   The asymptotic
nature of their treatment is nicely brought out by Van Kampen(19 ZZ) and by Dellar (19ww).

Let me mention in passing that dimensional reduction allows us to
pass from the  partial differential equations governing convection in a fluid layer to the sort of coupled ordinary differential discussed by Lorenz as a model displaying deterministic chaos.   Michel devised his famous attractor to model that chaotic behavior.  All this relates
to the present discussion and I will next try to provide an intuitive description of the underlying procedures for the the derivation of fluid dynamics.  For the purpose, I leave out the smoothed
phase space density so that ${\mathcal F}=f$ and that I need not rewrite (\ref{11})-(\ref{14})
for what follows. 

As I have expressed in words already, the basic conservation laws of mechanics may
be expressed here as (\ref{cons}).  When we apply that condition to \be
 \p_t f + {v^i \p_i} f + (\p_i \Phi) \p_{v^i}f = {\mathcal C}[f] .  \label{backB}\ee 
and carry out the integrations, we obtain  \begin{equation}
\partial_t \, \rho+\nabla\cdot (\rho {\bf u})=0 \label{ma}
\end{equation} \begin{equation}
 \partial_t {\bf u}+ {\bf u}\nabla {\bf u}- {\bf a}+\frac{1}{\rho}
\nabla\cdot{\mathbb P} =0 \label{mo}
\end{equation} \begin{equation}
 \partial_t T+{\bf u}\cdot\nabla T+\frac{2}{3\Re \rho}({\mathbb P}: 
\nabla {\bf u}+\nabla\cdot {\bf Q}) =0\, ,
\label{en}
\end{equation}
where the colon stands for a double dot product and ${\bf a}$ is given by (\ref{a})
in terms of $\Phi$ that we get from Poisson's equation, (\ref{rho}). 
 
Thus continuum equations are a formal consequence of the kinetic equation.  
However, these equations are not self-contained for we have not provided a prescription for determining $\mathbb P$ and ${\bf Q}$.  The challenge that remains is to complete the system of equations (\ref{ma})-(\ref{en}) with closure representations for the undetermined quantities so that we obtain a self-contained system of equations whose solutions lead to
a reasonably accurate representation of the behavior of the fluid in terms of the fluid variables 
$\rho, T$ and ${\bf u}$.  Following Hilbert's lead, we may do this by seeking an approximate    
of (\ref{backB}).  To indicate the direction in which to go I open a penultimate section.
\section{The End in Sight}
It is not very difficult to find suitable approximations for ${\mathbb P}$
and ${\bf Q}$ to close the system (\ref{ma})-(\ref{en}) by seeking an approximate solution
of (\ref{backB}).  The key is to recognize that the right hand side of $(\ref{backB})$ changes on the mean flight time of the constituent particle while the left hand side evolves on a characteristic macroscopic time scale, the so-called streaming time.  Their ratio, is known as the Knudsen number for a Danish physicist who studied rarefied gases.  For the propagation of sound waves, the relevant macroscopic time is the period.  We may introduce as dimensionless parameter, the mean free path times the acoustic wave number.  In that case, the dimensionless wavenumber $k$ is the Knudsen number.  So the problem of completing the fluid equations comes down to seeking an approximate solution of \be 
k{\mathcal D}f = {\mathcal C}[f] \ee
where \be
{\mathcal D} = \p_t  + {v^i \p_i}  + (\p_i \Phi) \p_{v^i} \ee
and where the collision model expressed in ${\mathcal C}[f]$ has to be chosen. 

Since $k$ appears in front of a derivative,
we have to deal with a problem in singular perturbation theory and so we must be cautious.
In leading (or zeroth) order we are led to the Euler equation for the gas dynamics. But in
the Chapman-Enskog approach one simplifies the next order by using the results of the
previous order.  This leads, in first order, to the Navier-Stokes equations.  And, in the following order, in using this iterative approach, one arrives at the Burnett equations,
which are less good than the N-S equation (Grad, 19??).  That is a sign that we are
dealing with an asymptotic sequence and not a convergent series.  This is not a good
place to iterate.  As Uhlenbeck has reported in his lectures (Uhlenbeck and Ford 19??),
the phase speed and the damping length of sound as predicted by the N-S equations
do not agree with the results of experiment (at least for $k>1$).  If one does not iterate,
one gets a more general equation for the gas dynamics and a reasonably good result for
the phase speed of sound waves.  The damping length of sound waves is not yet well
predicted and that story is too long for this account.  But stay tuned if this problem 
interests you.  

In any case, the secret is now out --- the fluid variables are slow
variables (at least in Boltzmann's model) in consequence of the conservation laws and that is why they control the destiny of the fluid.  That is the idea underlying center manifold theory.  Astrophysicists will recognize this as just the sort of idea they use to allow slowly reacting species to guide the evolution of stars.  I suspect may be some weak sense in which Uhlenbeck's worries may be resolved.   A proof of "Hilbert's theorem," which holds
under suitably restricted conditions, may yet be found, if it does not exist already.
\section{A Final Word}
Whoever said ``If I'd had more time, I would have written less" was speaking
for me on this occasion.  It would be insensitive to continue further but, if you are
still reading, please allow me one last remark.

Einstein said that ``things should be made as simple as possible, but no simpler."
That is the thinking of genius of course and few people have the gift of going right to
heart of a problem with no wasted motion.  Michel H\'enon had that gift and those of
us who have studied his works are keenly aware of this.  That someone of such 
manifest gifts should also behave in such an unpretentious manner made him an 
all the more admirable personage whose memory we rightly honor here.

I am grateful to Engelbert Schucking for a number of informative conversations and
also to the several friends too numerous to list here that patiently listened to my retelling of
this story.  I will tell it till I get it right.

\end{document}